\PassOptionsToPackage{bottom,flushmargin}{footmisc}
\PassOptionsToPackage{sort&compress}{natbib}

\documentclass[a4paper]{cas-dc}
\RenewDocumentCommand{\printorcid}{}{} 

\usepackage[numbers]{natbib}
\usepackage{graphicx}
\usepackage{amsmath,amssymb}
\usepackage{float}
\usepackage{xcolor}
\usepackage{bm}
\usepackage{mathtools}
\usepackage{tensor}
\usepackage{cancel}
\usepackage{physics}
\usepackage{caption}
\usepackage{framed}
\usepackage{multirow}
\usepackage{makecell}

\renewcommand{\H}{{\cal H}}
\renewcommand{\d}{\text{d}}

\newcommand{\orho}{\bar\rho}
\newcommand{\opsi}{\bar\psi}
\newcommand{\oxi}{\bar\xi}
\newcommand{\crit}{\rho_\text{crit}}
\newcommand{\DM}{\text{c}}
\newcommand{\DE}{\text{DE}}
\newcommand{\eff}{\text{eff}}
\newcommand{\B}{\text{B}}
\newcommand{\R}{\text{R}}
\newcommand{\D}{\text{D}}
\renewcommand{\c}{\text{c}}
\newcommand{\lsim}   {\mathrel{\mathop{\kern 0pt \rlap
  {\raise.2ex\hbox{$<$}}}
  \lower.9ex\hbox{\kern-.190em $\sim$}}}
\newcommand{\gsim}   {\mathrel{\mathop{\kern 0pt \rlap
  {\raise.2ex\hbox{$>$}}}
  \lower.9ex\hbox{\kern-.190em $\sim$}}}

\usepackage{eso-pic}
\newcommand{\PreprintNumber}{IPARCOS-UCM-26-039}

\begin{document}
\fontsize{9}{11}\selectfont

\title[mode = title]{The silent dark sector: a field-theory approach to unified dark fluids with vanishing speed of sound}

\author{Javier {de Cruz Pérez}}
\ead{j.decruz26@gmail.com}

\author{Darío Jaramillo-Garrido}
\cormark[1]
\ead{djaramil@ucm.es}

\author{Antonio L. Maroto}
\ead{maroto@ucm.es}

\author{Prado Martín-Moruno}
\ead{pradomm@ucm.es}

\cortext[cor1]{Corresponding author}

\affiliation{organization={Departamento de Física Teórica and Instituto de Física de Partículas y del Cosmos (IPARCOS-UCM)}, addressline={\\Universidad Complutense de Madrid},postcode={28040}, city={Madrid}, country={Spain.}}

\begin{abstract}
    We introduce a novel class of field-theory models that provide a unified description of the dark sector as a single perfect fluid with vanishing sound speed.  These models also admit the interpretation of cold dark matter interacting with vacuum energy while avoiding any phantom behavior and including standard $\Lambda$CDM as a particular case. Built upon a single scalar field invariant under transverse diffeomorphisms, the framework naturally implements the constraints imposed in the so called mimetic models. We analyze a particular simple example exhibiting an effectively evolving dark energy and find that it can reproduce the DESI best-fit CPL evolution with the same number of free parameters.
\end{abstract}


\AddToShipoutPictureFG*{%
  \AtPageUpperLeft{%
    \put(\LenToUnit{\paperwidth-20mm},%
         \LenToUnit{-10mm}){%
      \makebox(0,0)[r]{\small\PreprintNumber}%
    }%
  }%
}

\maketitle

\section{Introduction}
\noindent The lack of understanding regarding the fundamental nature of dark matter and dark energy, together with recent tensions between different cosmological datasets \cite{Riess:2021jrx,DESI:2025zgx}, has driven the search for descriptions of the dark sector beyond the standard $\Lambda$CDM framework \cite{CosmoVerseNetwork:2025alb}. The most widely considered alternatives include models where dark energy evolves in time \cite{deCruzPerez:2025dni}, theories with interacting dark matter and dark energy \cite{Wang:2016lxa,Wang:2024vmw,Li:2026xaz}, or models in which the complete dark sector is studied as a whole from a unified point of view \cite{Scherrer:2004au,Bertacca:2010ct}. These ideas can ultimately be motivated by the so-called ``dark degeneracy'' \cite{Wasserman:2002gb,Kunz:2007rk,Kunz:2009yx}, which alludes to the fact that cosmological observations are not sensitive to the individual components of the dark sector but rather to the total energy-momentum tensor, hence allowing us to split it in different ways.

Recent observations of baryon acoustic oscillations carried out by the Dark Energy Spectroscopic Instrument (DESI) combined with CMB and supernova data \cite{DESI:2025zgx} seem to indicate a certain preference for an evolving dark energy component, (see also the more recent, calibration-corrected \cite{Popovic:2025glk} results in \cite{DES:2025sig} and the discussions in e.g. \cite{Efstathiou:2024xcq,Nesseris:2025lke,Ong:2026tta} and references therein for criticisms). The analysis, based on the $w_0w_a$CDM or CPL (Chevallier-Polarski-Linder) parametrization \cite{Chevallier:2000qy,Linder:2002et}, also  favors a transition in the equation of state of dark energy from $w_\text{DE}<-1$ to $w_\text{DE}>-1$. However, this type of phantom crossing is known to pose consistency problems when trying to model dark energy from a field-theory point of view, including  the appearance of instabilities related to the violation of the null energy condition or the existence of singularities in the evolution of perturbations \cite{Caldwell:1999ew,Carroll:2003st,Vikman:2004dc,Hu:2004kh,Kunz:2006wc,BorislavovVasilev:2024loq}. Nonetheless, it has been recently proposed in \cite{Kou:2025yfr} that a unified description of the dark sector could provide a viable alternative to phantom dark energy. Thus, exploiting the dark sector degeneracy mentioned above, \cite{Kou:2025yfr} shows that by describing the total dark sector as a single perfect fluid with equation of state $w_\text{D}>-1$, it is possible to reproduce the background expansion without introducing any phantom component. At the perturbative level, in order to be compatible with structure formation, the speed of sound is assumed to be zero, thus satisfying the stringent limits imposed by the matter power spectrum observations \cite{Sandvik:2002jz} which typically set $\vert c_s^2\vert\lsim 10^{-5}$. The results presented
in \cite{Kou:2025yfr} show that this unified fluid description performs comparably to CPL when confronted with current data from DESI BAO DR2, Planck, and DESY5 and yields similar forecasts
for next-generation surveys. 
Given the purely phenomenological nature of that approach, establishing a fundamental model that provides a theoretical basis for such a unified fluid is of great interest and this is precisely the aim of the present work.
Different proposals have been considered in the literature for a unified description of the dark sector, either based on the generalized Chaplygin gas cosmologies or on $k$-essence models \cite{Scherrer:2004au,Bertacca:2010ct}. In the first case, the constraints imposed by the speed of sound make them practically indistinguishable from $\Lambda$CDM \cite{Sandvik:2002jz}, although a possible solution has been recently found by means of generalized Chaplygin-type solids \cite{BeltranJimenez:2026tlr}.  In the $k$-essence models, the matter power spectrum observations together with constraints from the integrated Sachs-Wolfe effect impose very stringent constraints on the model parameters \cite{Bertacca:2007cv}. More recent theoretical work combines the effective field theory of dark energy with open-system techniques to construct a unified dark sector \cite{Salcedo:2026cqb}.

A different approach to unification is built upon constrained dynamics, such as those of mimetic models \cite{Lim:2010yk,Sebastiani:2016ras}, which achieve a vanishing speed of sound  by forcing energy to flow along timelike geodesics. 
Although this condition ultimately leads to the unification of the dark sector, it must be imposed ad hoc at the Lagrangian level. However, \cite{deCruzPerez:2025ytd} recently showed that the mimetic constraint naturally arises in models invariant under transverse diffeomorphisms (TDiff) i.e. general coordinate transformation with unit Jacobian. In fact, a single TDiff scalar field with a purely kinetic term was shown to exactly reproduce the complete dark sector of $\Lambda$CDM in any background geometry.
In the present work, we build on this type of theories and identify the class of TDiff scalar field models which are able to unify cold dark matter (CDM) and  dynamical dark energy while maintaining a vanishing speed of sound.

The paper is organized as follows. In section \ref{section: Fundamental framework} we discuss the theoretical background, presenting the underlying field-theoretic framework and showing how to treat the field theory as a perfect fluid. In section \ref{section: Fluids with a vanishing speed of sound} we construct the possible vanishing sound speed models and show that they effectively implement the constraints imposed in the so-called mimetic theories. Section \ref{section: Dark sector applications} discusses the dark sector applications of the selected models, showing that they admit the interpretation of interacting vacuum models and providing a fundamental derivation of the interacting kernel. From that point onward, we therefore study the cosmological implications both at the background and perturbation levels, with the general treatment developed in section \ref{section: Cosmological dynamics} and carried out for a concrete quadratic potential in section \ref{section: Quadratic potential}, where we also compare with CPL models. Finally, section \ref{section: Conclusions} contains the summary and main conclusions. We will use throughout units such that $\hbar=c=1$, metric signature $(+,-,-,-)$, and denote $g=|\det(g_{\mu\nu})|$.


\section{Fundamental framework}\label{section: Fundamental framework}
\noindent With the aim of providing a more fundamental description of the unified dark fluid, in this section we begin by considering a canonical scalar field theory that breaks diffeomorphism invariance down to transverse diffeomorphisms (TDiff). We then show that this TDiff theory is actually  equivalent to a family of Diff theories featuring two scalar fields, one of which acts as a spectator. Finally, we present the effective fluid description of this framework and provide the relevant expressions to calculate the effective speed of sound of cosmological perturbations.

\subsection{Field theories}\label{section:scalar-vector}
\noindent We begin by considering a canonical scalar field theory that breaks diffeomorphism invariance down to transverse diffeomorphisms \cite{Maroto:2023toq,Jaramillo-Garrido:2023cor,Alonso-Lopez:2023hkx,Jaramillo-Garrido:2024tdv,deCruzPerez:2025ytd}
\begin{equation}\label{eq: manifestly TDiff action}
    S_\text{TDiff}[g_{\mu\nu},\psi] = \int \d^4x \big[ f_k(g) X - f_v(g) V(\psi) \big] ,
\end{equation}
where $f_k(g)$ and $f_v(g)$ are arbitrary functions of the absolute value of the metric determinant mediating the coupling to gravity, and
\begin{equation}\label{X}
    X = \frac{1}{2} \nabla_\mu\psi \nabla^\mu\psi
\end{equation}
denotes the kinetic term. The form of the potential $V(\psi)$ and the kinetic and potential coupling functions $f_k(g)$ and $f_v(g)$ will be considered generic for the moment. We will only assume $f_k(g)>0$ to avoid possible instabilities. Invariance under the complete group of diffeomorphisms would be recovered if $f_k(g) = f_v(g) = \sqrt{g}$, but the symmetry is generally broken down to the TDiff subgroup. Now, as discussed in \cite{Maroto:2023toq,Jaramillo-Garrido:2024tdv,Bello-Morales:2024vqk} (and references therein), it is possible to restore the symmetry under the full group of diffeomorphisms via the introduction of an additional vector field $A^\mu$ (\textit{à la} Stueckelberg), and construct the covariantized theory
\begin{equation}\label{eq: covariantized action}
    S_\text{cov}[g_{\mu\nu},\psi,A^\mu] = \int \d^4x \, \sqrt{g} \Big[ H_k(Y) X - H_v(Y) V(\psi) \Big] ,
\end{equation}
where $Y=\nabla_\mu A^\mu$ is the divergence of the new vector field, and the new coupling functions $H_a$ ($a=k,v$) are related to the old ones $f_a$ via
\begin{equation}
    H_a(Y) = Y f_a(Y^{-2}).
\end{equation}
This covariantized theory can be translated back to the TDiff theory by evaluating everything in the so-called ``TDiff frame'', which essentially boils down to the substitutions
\begin{equation}
    Y \rightarrow \frac{1}{\sqrt{g}} ,\quad H_a(Y) \rightarrow \frac{f_a(g)}{\sqrt{g}} ,\quad H_a'(Y) \rightarrow f_a(g) - 2g f_a'(g) ,
\end{equation}
(primes denote differentiation with respect to the argument unless otherwise stated) see \cite{Jaramillo-Garrido:2024tdv} for further details. In essence, then, the two theories are the same, simply expressed in different coordinates. Since the vector field appears only through a divergence, one may naturally wonder whether it may be replaced by a second scalar field instead. This is indeed possible, but some subtleties should be taken into account in order to properly recover all solutions of theory, and we direct the reader to \cite{Blas:2011ac} for a detailed treatment. For the case at hand, the essential idea may be understood as follows \cite{deCruzPerez:2025ytd}: we can rewrite the scalar-vector action \eqref{eq: covariantized action} in the form
\begin{equation}\label{eq: lagrange multiplier}
    S = \int \d^4 x\sqrt{g} \, \Big[ H_k(\phi) X - H_v(\phi) V(\psi) + \mu\left(\nabla_\alpha A^\alpha - \phi\right) \Big] ,
\end{equation}
with $\phi$ a scalar field and $\mu$ a Lagrange multiplier enforcing $\phi=\nabla_\alpha A^\alpha = Y$. Variation with respect to the vector field $A^\alpha$ yields $\nabla_\alpha \mu = 0$, implying a constant $\mu = \mu_0$. In light of this result, we may rewrite \eqref{eq: lagrange multiplier} up to a boundary term as
\begin{equation}\label{eq: parametrized family}
    S[\psi,\phi] = \int \d^4 x\sqrt{g} \, \Big[ H_k(\phi) X - H_v(\phi) V(\psi) -\mu_0 \phi \Big] .
\end{equation}
The above expression defines a parametrized family of actions, one for each value of $\mu_0$ in the original theory, which recovers identical equations of motion. In this interpretation $\mu_0$ represents a global degree of freedom, and $\phi$ is a spectator field. 
Thus the theory in general contains one propagating degree of freedom $\psi$ and a global degree of freedom $\mu_0$ \cite{BeltranJimenez:2025pho}. However, as we will see in section \ref{section: Fluids with a vanishing speed of sound} for models with a vanishing speed of sound, the character of these fields will change.
The formulation \eqref{eq: parametrized family} with two scalar fields is more convenient to perform the computations, and will also give us a clear interpretation of the underlying dynamics later on. The equations of motion for the theory \eqref{eq: parametrized family} read
\begin{align}
    \delta\psi&:\quad \nabla_\alpha\big[H_k(\phi)\nabla^\alpha\psi\big] + H_v(\phi) V'(\psi) = 0 , \label{eq: scalar scalar KG}\\[5pt]
    \delta\phi&:\quad H_k'(\phi) X - H_v'(\phi) V(\psi) = \mu_0 , \label{eq: scalar scalar constraint}
\end{align}
and equation \eqref{eq: scalar scalar constraint} will be referred to as the ``constraint'' in the following. Finally, the associated energy-momentum tensor (EMT) takes the form
\begin{align}
    T_{\mu\nu} &= \frac{2}{\sqrt{g}} \frac{\delta S}{\delta g^{\mu\nu}} = \label{eq: scalar scalar EMT} \\[5pt]
    &= H_k(\phi) \nabla_\mu\psi \nabla_\nu\psi - \Big[ H_k(\phi) X - H_v(\phi) V(\psi) - \mu_0\phi \Big] g_{\mu\nu} . \nonumber
\end{align}
We now discuss the perfect fluid representation of the theory.

\subsection{The perfect fluid}\label{section: perfect fluid}
\noindent Under the assumption of the field derivative $\nabla_\mu\psi$ being a timelike ($X>0$) and future-pointing vector, we can write the EMT \eqref{eq: scalar scalar EMT} in perfect fluid form
\begin{equation}
    T_{\mu\nu} = (\rho + p) u_\mu u_\nu - p g_{\mu\nu} ,
\end{equation}
where the fluid velocity is defined as
\begin{equation}\label{eq: velocity definition}
    u_\mu = \frac{\nabla_\mu\psi}{\sqrt{2 X}} ,
\end{equation}
and we have the following energy density and pressure:
\begin{subequations}\label{eq: density and pressure}
    \begin{align}
        \rho &= H_k(\phi) X + H_v(\phi) V(\psi) + \mu_0 \phi ,\label{rho}\\[5pt]
        p &= H_k(\phi) X - H_v(\phi) V(\psi) - \mu_0 \phi .\label{p}
\end{align}
\end{subequations}
The equation of state (EoS) parameter is defined in the usual manner as $w=p/\rho$, and we now have all the fundamental background fluid quantities accounted for. Note that this fluid will satisfy the null energy condition
\begin{equation}\label{eq: rho + p > 0}
    \rho + p = 2 H_k(\phi) X \geq 0 
\end{equation}
whenever $H_k(\phi) \geq 0$, which is automatically satisfied based on our original assumption of $f_k(g)>0$.

Let us now move over to the discussion of the linear perturbations of these fluids. From the constraint equation \eqref{eq: scalar scalar constraint} it is possible to find that\footnote{We shall all throughout this work assume $H_k'(\phi)\neq 0$. The interested reader may find a discussion on the $H_k(\phi) = \text{const.}$ model in \cite{Jaramillo-Garrido:2024tdv}.}
\begin{equation}\label{Xconst}
    X = \frac{H_v'(\phi) V(\psi) +\mu_0}{H_k'(\phi)} = X(\psi,\phi) .
\end{equation}
This means that the energy density and pressure are functions of two variables $(\psi,\phi)$, and the general pressure perturbation will therefore be non-adiabatic. As discussed in \cite{Jaramillo-Garrido:2024tdv}, it takes the form
\begin{equation}\label{eq: delta p for field theory}
    \delta p = c_s^2 \delta\rho + \alpha \delta\psi,
\end{equation}
where we have defined
\begin{subequations}
    \begin{align}
        c_s^2 &= \frac{\partial p / \partial \phi}{\partial\rho / \partial \phi} , \label{eq: cs2 definition}\\[5pt]
        \alpha &= (-c_s^2) \frac{\partial \rho }{\partial\psi} + \frac{\partial p}{\partial\psi} .
    \end{align}
\end{subequations}
Note that $c_s^2$ will play the role of the effective speed of sound and $\alpha$ will be called the non-adiabatic coefficient. If $\alpha = 0$, we immediately have an adiabatic fluid satisfying $\delta p = c_s^2 \delta\rho$ in every frame. In addition, the expressions for $c_s^2$ and $\alpha$ in terms of the coupling functions can be obtained using \eqref{eq: density and pressure}. Momentarily suppressing the dependencies for clarity, these read
\begin{subequations}
    \begin{align}
        c_s^2 &= \frac{A}{A - B} \,, \label{cs2} \\[5pt]
        \alpha  &= \left[-c_s^2 \left( \frac{H_k}{H_k'} H_v' + H_v \right) +\left( \frac{H_k}{H_k'} H_v' - H_v \right)\right] V', \label{alpha}
    \end{align}
\end{subequations}
where
\begin{subequations}
    \begin{align} 
        A &= H_k \Big[ V \left( H_k'' H_v' - H_k' H_v'' \right) + \mu_0 H_k'' \Big] \,, \label{eq: A(Y,psi)}\\[5pt]
        B &= 2(H_k')^3 X\,,\label{eq: B(Y,psi)}
    \end{align}
\end{subequations}
as defined in \cite{Jaramillo-Garrido:2024tdv}. As a final small comment regarding perturbations, we draw attention to the fact that since EMT \eqref{eq: scalar scalar EMT} can be exactly written in perfect fluid form, neither heat flow nor anisotropic stress will arise when we perturb it.


\section{Fluids with a vanishing speed of sound}\label{section: Fluids with a vanishing speed of sound}
\noindent Now that the general framework for our unified dark sector models has been laid out, we restrict our focus to those models that are naturally compatible with structure formation, i.e. those characterized by a vanishing speed of sound for cosmological perturbations.
In this section, we will show that imposing the requirement of a vanishing sound speed fixes the coupling functions and, as a consequence, after suitable field redefinitions, action \eqref{eq: parametrized family} can be rewritten in such cases as
\begin{equation}\label{eq: mimetic action}
    S_{c_s^2=0}[\psi,\varphi] = \int \d^4 x\sqrt{g} \, \Big\{ X - V(\psi) + \varphi \left[X - \beta V(\psi) - \gamma \right] \Big\} ,
\end{equation}
where $\beta$ and $\gamma$ are two free parameters. Thus, we see that the action agrees with those in \cite{Lim:2010yk}, i.e.  Diff breaking realizes in a natural way the idea behind mimetic models  (see also the recent discussion in \cite{BeltranJimenez:2025pho}\footnote{Indeed, for $H_k'\neq0$, the condition for a vanishing speed of sound, $A=0$ with $A$ defined in equation \eqref{eq: A(Y,psi)}, is just the condition found in \cite{BeltranJimenez:2025pho} for having equivalence between the TDiff and the mimetic framework.}). We will also see that the choices $\beta=1$ and/or a constant $V(\psi) = V_0$ result in adiabatic fluids.

\subsection{General solutions}\label{section: mimetic}
\noindent In the previous section the form of the intervening functions was kept arbitrary, describing therefore a large family of models with different characteristics. For example, it was recently found in reference \cite{deCruzPerez:2025ytd} that a shift-symmetric theory with linear coupling $H_k(\phi) = k_1\phi + k_2$ results in an exactly vanishing sound speed and leads to a perfect match with the $\Lambda$CDM dark sector. Following this spirit, let us look for all possible vanishing sound speed models. In view of equation \eqref{cs2}, it is possible to see that $c_s^2=0$ amounts to requiring $A=0$ while keeping $B\neq 0$.\footnote{Indeed, if $B$ vanished the sound speed would be forced to take on the value $c_s^2=1$.} On the one hand, the non-vanishing of $B$ is essentially the non-vanishing of $X$ as follows from \eqref{eq: B(Y,psi)}, which means we must always make sure to satisfy $X\neq0$. On the other hand, from \eqref{eq: A(Y,psi)} the vanishing of $A$ can be written as
\begin{equation}\label{eq: vanishing condition}
   V(\psi) \, \left[H_k'(\phi)\right]^2 \,\frac{\d}{\d \phi}\left(\frac{H_v'(\phi)}{H_k'(\phi)}\right) = \mu_0 H_k''(\phi) .
\end{equation}
We now look for solutions to this equation.

\medskip
\noindent\textbf{Solution 1.} A linear kinetic coupling,
\begin{equation}
    H_k(\phi) = k_1 \phi + k_2 ,
\end{equation}
yields $H_k''(\phi)= 0$, which upon substitution in \eqref{eq: vanishing condition} implies that one can have $V(\psi)=0$, or
\begin{equation}
    \frac{\d}{\d \phi}\left(\frac{H_v'(\phi)}{H_k'(\phi)}\right) = 0\,\implies\, H_v(\phi) = c_1 H_k(\phi) +c_2
\end{equation}
for an arbitrary potential. Notice that we can, without loss of generality, demand that $c_1$ and $c_2$ do not vanish at the same time.\footnote{Otherwise $H_v(\phi)=0$, which is degenerate with the $V(\psi)=0$ scenario (they represent the shift-symmetric theory).} Substituting these coupling functions in our (parametrized) action \eqref{eq: parametrized family} gives a convoluted expression, but it is possible to simplify it as follows. Firstly, we define a new field $\varphi$ through
\begin{equation}\label{eq: def varphi}
    \varphi = \frac{H_k(\phi)}{\sigma^2} - 1 ,
\end{equation}
with $\sigma^2>0$ a positive constant introduced such that $c_1\sigma^2+c_2\neq0$ (for reasons that will become apparent in a moment), but arbitrary beyond that. Secondly, we perform the substitutions
\begin{equation}\label{eq: field redefinition}
    \psi \ \to \ \frac\psi\sigma ,\qquad V(\psi) \ \to \ \frac{V(\psi) - \mu_0 (\sigma^2 - k_2)/k_1}{c_1\sigma^2 + c_2}
\end{equation}
which is of course only well defined for $c_1\sigma^2+c_2\neq0$, hence our previous requirement. Finally, we define the constants
\begin{align}
    \beta &= \frac{c_1 \sigma^2}{c_1\sigma^2 + c_2} , \label{eq: def beta}\\[5pt]
    \gamma &= \frac{\mu_0\sigma^2(c_1 k_2 + c_2)}{k_1 (c_1\sigma^2 + c_2)}. \label{eq: def gamma}
\end{align}
Using all of the above, we may bring \eqref{eq: parametrized family} into the form
\begin{equation}\label{eq: action for sol 1}
    S = \int \d^4 x\sqrt{g} \, \Big\{ X - V(\psi) + \varphi \left[X - \beta V(\psi) - \gamma \right] \Big\},
\end{equation}
where we immediately notice that the new field $\varphi$ plays the role of a Lagrange multiplier enforcing a relation between the kinetic term $X$ and the potential $V(\psi)$. The particular case where $V(\psi)=0$ is also accommodated in this framework, and reproduces the results in \cite{deCruzPerez:2025ytd} for the shift-symmetric case. Note that in general the Lagrange multiplier enforces $X = \beta V(\psi) + \gamma$, and recalling the restriction $X\neq 0$ we conclude that the two parameters $\beta$ and $\gamma$ cannot vanish at the same time.

\medskip
\noindent\textbf{Solution 2.} Whenever $H_k''(\phi)\neq0$, the right-hand side of \eqref{eq: vanishing condition} does not automatically vanish. It does however if we assume $\mu_0=0$, from which \eqref{eq: vanishing condition} implies we can have $V(\psi)=0$ or 
\begin{equation}
    H_v(\phi) = c_1 H_k(\phi) + c_2 .
\end{equation}
Both situations are accommodated substituting in \eqref{eq: parametrized family} to yield
\begin{equation}\label{eq: action with arbitrary V}
    S = \int \d^4 x\sqrt{g} \, \Big\{ H_k(\phi) \big[X - c_1 V(\psi)\big] - c_2 V(\psi) \Big\} .
\end{equation}
Notice however that the shift symmetric situation (defined by $c_1=c_2=0$ and/or $V(\psi)=0$) is forbidden as it would yield $X=0$ from \eqref{eq: scalar scalar constraint} and this is not allowed. Proceeding now analogously to how we did for Solution 1, we introduce a field $\varphi$ through \eqref{eq: def varphi}, perform the same redefinitions as in \eqref{eq: field redefinition} (recalling $\mu_0=0$), and define the same $\beta$ as in \eqref{eq: def beta}, to rewrite action \eqref{eq: action with arbitrary V} as
\begin{equation}\label{eq: action for sol 2}
    S = \int \d^4 x\sqrt{g} \, \Big\{ X - V(\psi) + \varphi \left[X - \beta V(\psi)\right] \Big\}.
\end{equation}
Thus, the dynamics will be given by the same theory found for Solution 1 in \eqref{eq: action for sol 1}, only now with one of the parameters fixed to vanish ($\gamma=0$).

\medskip
\noindent\textbf{Solution 3.} Finally, consider the situation in which $H_k''(\phi)\neq 0$ and $\mu_0\neq0$. In such a case, the right-hand side of \eqref{eq: vanishing condition} does not vanish and is instead a general function of $\phi$. In order not to obtain a relation between $\psi$ and $\phi$ (which would be a particular solution), we must impose a constant potential $V(\psi)=V_0\neq 0$. We then use \eqref{eq: vanishing condition} to solve for the potential coupling function, yielding
\begin{equation}
    H_v(\phi) = c_1 H_k(\phi) + c_2 - \frac{\mu_0}{V_0} \phi .
\end{equation}
Substituting everything in \eqref{eq: parametrized family} gives
\begin{equation}\label{eq: action for sol 3}
    S = \int \d^4 x\sqrt{g} \, \Big[ H_k(\phi) \big(X - c_1 V_0\big) - c_2 V_0 \Big] .
\end{equation}
It is immediate to notice that \eqref{eq: action for sol 3} is precisely the previous \eqref{eq: action with arbitrary V} evaluated for the constant potential $V(\psi)=V_0\neq0$ (in particular, we can again discard the shift-symmetric case $c_1=c_2=0$, as it would imply the forbidden $X=0$). An identical procedure then shows that we may write it in the form \eqref{eq: action for sol 2} with analogous definitions and rescalings, only now writing a constant $V_0$ in place of the previously arbitrary $V(\psi)$.

\medskip
\noindent\textbf{Summary.} We thus conclude that all vanishing sound speed models can be treated within the common framework provided by the mimetic theory \eqref{eq: mimetic action}, as we advanced at the beginning of the section. This shows that single-field TDiff models with vanishing sound speed in general have two free parameters, $\beta$ and $\gamma$. The equations of motion which follow from \eqref{eq: mimetic action} read
\begin{align}
    \delta\psi&:\quad \nabla_\alpha\big[(1+\varphi)\nabla^\alpha\psi\big] + (1+\beta\varphi) V'(\psi) = 0, \label{eq: mimeticfield}\\[5pt]
    \delta\varphi&:\quad X = \beta V(\psi) + \gamma . \label{eq: Xconst}
\end{align}
Note that, focusing in $c_s^2=0$, the original dynamical system consisting of one second order differential equation for $\psi$ \eqref{eq: scalar scalar KG} and a constraint for $\phi$ \eqref{eq: scalar scalar constraint} has now become a system of two first order differential equations, equations \eqref{eq: mimeticfield} and \eqref{eq: Xconst}. Thus, no field has a wavelike equation in this case (see \cite{Lim:2010yk} for further details).
Finally, the EMT can be expressed as
\begin{equation}\label{eq: EMT mimetic}
    T_{\mu\nu} = (1+\varphi) \nabla_\mu\psi \nabla_\nu\psi - \big[ (\beta-1) V(\psi) + \gamma \big] g_{\mu\nu} ,
\end{equation}
where it is interesting to note for future purposes that $\varphi$ appears now only in the first term.

\subsection{Adiabatic models}\label{section: adiabaticity}
\noindent Although in general the models introduced above describe non-adiabatic fluids, some particular cases correspond in fact to adiabatic fluids. They are characterized by the vanishing of the non-adiabatic coefficient $\alpha$ introduced in \eqref{alpha}. As we are selecting models in which $c_s^2=0$, we focus on
\begin{equation}
    \alpha\big|_{c_s^2=0} = V'(\psi) \, \frac{H_k(\phi)^2}{H_k'(\phi)} \, \frac{\d}{\d \phi}\left( \frac{H_v(\phi)}{H_k(\phi)} \right) = 0 .
\end{equation}
This is achieved for a constant potential and/or for proportional couplings $H_v(\phi) = c_1 H_k(\phi)$. We can easily translate this to the language of the mimetic theory \eqref{eq: mimetic action} by noting that, on the one hand, rescaling a constant potential is just another constant potential, which makes Solution 3 adiabatic by default. Solutions 1 and 2 have an additional way of being adiabatic, revealed by noting that proportional couplings have $c_2=0$ and so $\beta=1$ as follows from \eqref{eq: def beta}. We thus conclude that the adiabatic models with vanishing sound speed are quite simply characterized by $V(\psi)=V_0$ and/or $\beta=1$. In any other situation, we shall be dealing with a non-adiabatic fluid.


\section{The unified fluid as an interacting dark sector}\label{section: Dark sector applications}
\noindent In order to understand the nature of the vanishing sound speed  models given by the action in \eqref{eq: mimetic action}, we note that the EMT \eqref{eq: EMT mimetic} can be written in perfect fluid form with velocity \eqref{eq: velocity definition} and an energy density and pressure which we suggestively denote as follows:
\begin{align}
    \rho_\D &= \underbrace{2(1+\varphi) \big[\beta V(\psi) + \gamma\big]}_{\equiv\,\rho_\c} + \underbrace{ (1-\beta) V(\psi) - \gamma }_{\equiv\,\rho_\lambda} , \label{eq: rho decomposition}\\[5pt]
    p_\D &= \underbrace{ (\beta-1) V(\psi) + \gamma }_{\equiv\,p_\lambda}. \label{eq: p decomposition}
\end{align}
The above decomposition makes explicit that all such models admit an interpretation as the sum of two comoving perfect fluids: a dust component with energy density $\rho_\c$ and pressure $p_\c=0$ (hence $w_\c=0$), and a vacuum energy component with energy density $\rho_\lambda$ and pressure $p_\lambda = -\rho_\lambda$ (hence $w_\lambda = -1$ exactly). The complete EMT can thus be generally decomposed \cite{Wands:2012vg} as
\begin{equation}\label{eq: EMT decomposition}
    T_{\mu\nu}^{(\D)} = T^{(\c)}_{\mu\nu} + T^{(\lambda)}_{\mu\nu} ,
\end{equation}
where each piece is of perfect fluid form. It is worth noting that since the two components are not separately conserved, these models  effectively act as interacting vacuum models \cite{Wands:2012vg,Wang:2013qy,Wang:2014xca,Salvatelli:2014zta} (we also direct the interested reader to literature on running vacuum models, e.g. \cite{Sola:2014tta,Sola:2015wwa,Sola:2016jky,SolaPeracaula:2016qlq}).

Further considering the interacting dark sector interpretation (see e.g. \cite{Wang:2016lxa,Wang:2024vmw} for reviews), let us compute the so-called interacting kernel $Q^\nu$. This quantity, defined through
\begin{equation}
    Q^\nu = \nabla_\mu T_{(\c)}^{\mu\nu} = - \nabla_\mu T_{(\lambda)}^{\mu\nu} ,
\end{equation}
is a measure of the energy-momentum exchange. We can gain some more intuition as to what are its effects by projecting the dust contribution's EMT conservation equation onto the direction parallel to the fluid velocity:
\begin{equation}
    u_\nu Q^\nu = u_\nu \nabla_\mu T^{\mu\nu}_{(\c)} = u^\nu\nabla_\nu \rho_\c + \rho_\c \nabla_\nu u^\nu .
\end{equation}
Denoting the projected derivative as $u^\nu \nabla_\nu \rho_\c = \dot\rho_\c$, the expansion as $\nabla_\nu u^\nu= \Theta$, the projected kernel as $u_\nu Q^\nu = Q$, and rearranging, it follows
\begin{equation}
    \dot\rho_\c = - \rho_\c \Theta + Q .
\end{equation}
The first piece in the right-hand side has the usual interpretation: the energy density of dust dilutes if the expansion $\Theta$ is positive and grows if it is negative. The interpretation of the second piece is just as simple: the dust component will receive an external energy if $Q>0$, while it will lose energy if $Q<0$. In cases (like the ones presented here) where the DM and DE components are comoving, the projection $Q$ is exactly the DM/DE energy transfer.\footnote{Some phenomenological interacting DM/DE models allow for the possibility that the two components of the dark sector move at different velocities, and so $Q^\nu \propto (u_\text{DM}^\nu - u_\text{DE}^\nu)$, see e.g. \cite{Asghari:2019qld,BeltranJimenez:2020iyx}. In a cosmological context, this is then only observed for the perturbations since the background velocity is assumed to be shared by all components.} For the models under study, the interacting kernel reads
\begin{equation}\label{eq: interacting kernel}
    Q^\nu = \nabla^\nu p_\lambda = (\beta-1) V'(\psi) \nabla^\nu\psi ,
\end{equation}
which is actually directed parallel to the velocity $u^\nu\propto\nabla^\nu\psi$.
Note that this is ultimately a consequence of the vacuum energy being a function only of $\psi$.
Incidentally, the fact that the interacting kernel is proportional to a scalar field gradient is reminiscent of coupled quintessence \cite{Amendola:1999er}. The projection onto the velocity is also straightforward,
\begin{equation}\label{eq: projected kernel Q}
    Q = (\beta-1)V'(\psi) \sqrt{2X} ,
\end{equation}
and because the square root is positive we conclude that energy flows from the vacuum energy to the dust component if $(\beta-1) V'(\psi)>0$, and the other way around if $(\beta-1)V'(\psi)<0$. It is interesting to note that there are many different ansätze for the interacting kernel in the cosmological literature, usually phenomenological in nature (see e.g. \cite{Wang:2013qy}).
In the present work, 
the underlying field theory provides an explicit expression for the interacting kernel.

\medskip
\noindent\textbf{Recovering standard cosmology.} We close this section by noting that the (non-)adiabaticity of these particular models also plays a crucial role in the dark sector interactions within this scenario. Indeed, recall that the adiabatic models were defined by $V(\psi)=V_0$ and/or $\beta = 1$. In any of these cases we find no energy exchange as the interacting kernel \eqref{eq: interacting kernel} vanishes. But this was to be expected, because for these adiabatic models the vacuum energy contribution introduced in \eqref{eq: rho decomposition} takes on a constant value $\rho_\lambda = \lambda_0$, and the total energy density and pressure \eqref{eq: rho decomposition} and \eqref{eq: p decomposition} respectively become
\begin{align}
    \rho_\D &= \rho_\c + \lambda_0 ,\\[5pt]
    p_\D &= -\lambda_0.
\end{align}
We therefore conclude that adiabatic models exactly reproduce the $\Lambda$CDM dark sector. Moreover, because we have performed a covariant analysis, this identity holds exactly in any spacetime. This extends the treatment carried out in \cite{deCruzPerez:2025ytd}, where it was shown that one could describe $\Lambda$CDM via a shift-symmetric vanishing sound speed model of the type here discussed (which may now be understood as choosing $V(\psi)=0$ in Solution 1 obtained in section \ref{section: mimetic}). Note finally that it is possible to make the vacuum energy contribution disappear: the choice $\beta=1$ and $\gamma=0$ yields $\rho_\lambda=0$, which would be a very simple dark matter model.


\section{Cosmological dynamics}\label{section: Cosmological dynamics}
\noindent In this section we shall consider the cosmological dynamics of the selected models. We will obtain
the general evolution equations both at the background and perturbation levels.  

\subsection{Cosmological background}
\noindent Let us choose a spatially flat FLRW spacetime as the background,
\begin{equation}
    \d s^2 = \d t^2 - a^2(t) \d\textbf{x}^2 ,
\end{equation}
with $a(t)$ the scale factor. As usual, we will assume that the only dependence of background quantities is in the cosmic time $t$, and a dot will denote differentiation with respect to it ($\,\dot{ \ }=\d/\d t$).
We shall consider a universe filled with baryons (B), radiation (R), and our vanishing sound speed fluid accounting for the dark sector (D). The Friedmann and acceleration equations respectively read
\begin{align}
    H^2 &= \frac{\kappa^2}{3} \left( \rho_\D + \rho_\B + \rho_\R \right) , \label{eq: Friedmann} \\[5pt]
    \dot H &= - \frac{\kappa^2}{2} \left( \rho_\D + p_\D + \rho_\B + \frac43 \rho_\R\right) ,
\end{align}
where $\kappa^2 = 8\pi G$, and $H=\dot a / a$ is the Hubble rate. Recalling from \eqref{eq: rho + p > 0} that $\rho_\D + p_\D \geq 0$, we see that $\dot H \leq 0$; so, the cosmic expansion cannot accelerate more than that of a de Sitter model. The evolution of the energy densities corresponding to baryons and radiation follows from their individual energy conservation equations,
\begin{align}
    \rho_\B &= \rho_{\B0} \,a^{-3} ,\\[5pt]
    \rho_\R &= \rho_{\R0} \, a^{-4} ,
\end{align}
with $\rho_{\B0},\,\rho_{\R0}$ their present day values. For the evolution of the dark sector we will, on the one hand, have the constraint \eqref{eq: Xconst}
\begin{equation}\label{Xcosmo}
    X = \frac12 \dot\psi^2 = \beta V(\psi) + \gamma ,
\end{equation}
which yields
\begin{equation}\label{eq: psi dot}
    \dot\psi = \sqrt{2\big[ \beta V(\psi) + \gamma \big]} ,
\end{equation}
and, on the other hand, the dark fluid conservation equation
\begin{equation}
    \dot\rho_\D + 3H (\rho_\D + p_\D) = 0 .
\end{equation}
Using the decomposition in  \eqref{eq: rho decomposition} and \eqref{eq: p decomposition} we rewrite the above equation as
\begin{equation}\label{eq: rhoc dot}
    \dot\rho_\c + 3H\rho_c = -\dot\rho_\lambda .
\end{equation}
The EoS parameter for the dark sector $w_\D = p_\D / \rho_\D$ takes the simple form
\begin{equation}\label{eq: DS EoS}
    w_\D = \frac{-\rho_\lambda}{\rho_c + \rho_\lambda} .
\end{equation}
We can in general choose the dynamical variables for the cosmological background problem to be the scalar field $\psi$, the dust energy density $\rho_\c$, and the Hubble rate $H$. In terms of these variables, the equations \eqref{eq: psi dot}, \eqref{eq: rhoc dot}, and \eqref{eq: Friedmann} respectively become
\begin{align}
    \frac{\d\psi}{\d a} &= \frac1{aH} \sqrt{2\big[ \beta V(\psi) + \gamma \big]} , \label{eq: dpsi/da}\\[5pt]
    \frac{\d\rho_c}{\d a} &= - \frac3a \rho_c - (1-\beta) V'(\psi) \frac{\d \psi}{\d a} , \label{eq: drhoc/da} \\[5pt]
    H^2 &= \frac{\kappa^2}{3} \Big[ \rho_\c + (1-\beta) V(\psi) - \gamma +\rho_{\B0}\, a^{-3} + \rho_{\R0}\, a^{-4} \Big] \label{eq: Friedmann final} ,
\end{align}
where we have changed the independent variable from cosmological time to the scale factor recalling that $\d a = aH \d t$. Given a potential $V(\psi)$, these are the equations that determine the background evolution.

As a quick consistency test let us consider the adiabatic models, characterized by $V(\psi)=V_0$ and/or $\beta = 1$. Both options result in the scaling $\rho_\c \propto a^{-3}$ from the conservation equation \eqref{eq: drhoc/da}, as befits a (non-interacting) cold dark matter component. We thus verify that the $\Lambda$CDM limit is explicitly recovered as expected.

\subsection{Cosmological perturbations}
\noindent Here we shall first review the general theory of cosmological perturbations for a perfect fluid (see e.g. \cite{Kodama:1984ziu,Ma:1995ey,Mukhanov:2005sc,Baumann:2022mni}) and then apply it to the vanishing sound speed dark sector that we have constructed. We shall only consider scalar perturbations and work, for simplicity, in the longitudinal gauge. In the end, we will also express the resulting equations in the synchronous gauge as it can yield convenient simplifications in some cases.

\medskip
\noindent\textbf{Metric and matter perturbations.} In the longitudinal gauge the scalar-perturbed FLRW metric reads
\begin{equation}
    \d s^2 =a^2(\eta)\left[(1+2\Phi)\,\d\eta^2-(1-2\Psi)\,\d\textbf{x}^2\right] ,
\end{equation}
with conformal time $\eta$ defined through $\d\eta = \d t/a(t)$. For simplicity, a prime in this subsection will denote derivative with respect to conformal time ($\, '=\d/\d\eta$). For the material content, one needs the velocity perturbation $v^i = \d x^i/\d\eta$, defined through
\begin{equation}
    u^\mu = u^\mu_0 + \delta u^\mu = \frac{1}{a} \left(1-\Phi,\,v^i\right) ,
\end{equation}
where a subindex ``$0$'' will denote background quantities whenever we are in the context of perturbations. Because our models define the fluid velocity through \eqref{eq: velocity definition} and because $\psi(\eta,\textbf{x}) = \psi_0(\eta) + \delta\psi(\eta,\textbf{x})$, the velocity perturbation has no direct vector contribution and is instead purely determined by the velocity potential
\begin{equation}\label{eq: velocity potential}
    v=-\frac{\delta\psi}{\psi_0'} 
\end{equation}
according to $v_i = \delta_{ij} v^j = v,_i$.

\medskip
\noindent\textbf{Evolution equations.} On the one hand, the perturbed Einstein equations $\delta G\indices{^\mu_\nu} = \kappa^2 \, \delta T\indices{^\mu_\nu}$ read, in Fourier space,
\begin{align}
    -3 \H^2 \Phi - 3\H \Phi' - k^2\Phi &= \frac{\kappa^2}2 a^2 \delta\rho , \label{eq: einstein with delta rho} \\[5pt]
    \H\Phi + \Phi' &= -\frac{\kappa^2}2 a^2 (\rho_0 + p_0) v, \label{eq: einstein with v}  \\[5pt]
    2\H' \Phi + \H^2 \Phi + \Phi'' + 3\H\Phi' &= \frac{\kappa^2}2 a^2 \delta p , \label{eq: einstein with delta p}
\end{align}
where we denote the conformal Hubble rate as $\H=a'/a = aH$. On the other, EMT conservation yields the continuity and Euler equations, which in Fourier space respectively read:
\begin{align}
    \delta'&=-(1+w)(\theta-3\Psi')-3{\cal H} \left(\frac{\delta p}{\rho_0}-w\,\delta\right) ,\label{encons}\\[10pt]
    \theta'&= -\H \,(1-3\,c_a^2)\theta + \frac{k^2}{1+w}\frac{\delta p}{\rho_0} + k^2\Phi , \label{euler}
\end{align}
where $\delta=\delta\rho/\rho_0$, $\theta=-k^2v$ and $c_a^2 = p_0'/\rho_0'$; so, using the background energy conservation, the adiabatic speed of sound can be expressed as
\begin{equation}
    c_a^2 = w - \frac{w'}{3\H(1+w)} .  \label{ca}    
\end{equation}
The pressure perturbation for a single perfect fluid may generally written as \cite{Bean:2003fb,Kunz:2006wc,Valiviita:2008iv,Ballesteros:2010ks}
\begin{equation}\label{eq: long pressure perturbation}
    \delta p = c_s^2 \delta\rho + 3\H \left(c_s^2-c_a^2\right) \rho_0 (1+w) \frac{\theta}{k^2} ,
\end{equation}
where $c_s^2$ denotes the effective speed of sound, which is constructed in the fluid's rest frame (denoted with a little hat and defined by $\hat\theta = 0$) as
\begin{equation}\label{eq: rest frame sound speed}
    c_s^2 = \frac{\widehat{\delta p}}{\widehat{\delta\rho}} .
\end{equation}
Plugging \eqref{eq: long pressure perturbation} into the continuity and Euler equations \eqref{encons} and \eqref{euler}, we obtain the final set of equations
\begin{align}
   \delta' &= -(1+w)\left(\theta-3\Psi'\right)-3\H
   \left(c_s^2-w\right)\delta \nonumber\\[5pt]
   &\qquad\quad - 9\H^2 \left(c_s^2-c_a^2\right) (1+w) \frac{\theta}{k^2} , \label{enconsc}\\[5pt]
   \theta' &= -\H \left(1-3c_s^2\right) \theta + \frac{k^2c_s^2}{1+w} \delta + k^2\Phi , \label{eulersc}
\end{align}
valid for any non-interacting perfect fluid.

\medskip
\noindent\textbf{Vanishing sound speed.} We now apply these general expressions to our vanishing sound speed dark fluid. In order to avoid cluttering the notation further, we do not explicitly include a subindex ``$\D$'' in the following quantities; the reader must in the following understand that every quantity refers to the dark fluid. The first point we can observe is that the quantity which we introduced as $c_s^2$ back in equation \eqref{eq: delta p for field theory} is indeed the effective speed of sound in the rest frame since, according to its definition \eqref{eq: cs2 definition}, it is calculated on the hypersurfaces of constant $\psi$. The main simplification in our models comes from having precisely a vanishing speed of sound in the rest frame, which reduces the general pressure perturbation \eqref{eq: long pressure perturbation} down to
\begin{equation}\label{eq: short pressure perturbation}
    \delta p = - 3\H c_a^2 \rho_0 (1+w)\frac{\theta}{k^2} ,
\end{equation}
and as a result the continuity and Euler equations \eqref{enconsc} and \eqref{eulersc} simplify correspondingly:
\begin{align}
   \delta' &= - (1+w) \left(\theta-3\Psi'\right) + 3 \H w \delta + 9 \H^2 c_a^2 (1+w)\frac{\theta}{k^2} ,\label{encons dark} \\[5pt]
   \theta' &= -\H\theta + k^2\Phi.\label{eulersc dark}
\end{align}
These are the general evolution equations for the perturbations of the vanishing sound speed dark sectors here constructed. Note that the adiabatic models have $c_a^2=0$, and hence we indeed recover the standard $\Lambda$CDM expressions.

\medskip
\noindent\textbf{Synchronous gauge.} For the sake of completeness, we also provide the perturbation evolution equations in the synchronous gauge, as they may prove useful for future studies. In this case the scalar-perturbed FLRW metric reads
\begin{equation}
    \d s^2 =a^2(\eta) \left\{ \d\eta^2-\big[(1-2\Psi)\delta_{ij} + 2 E,_{ij}\big] \d x^i \d x^j \right\} .
\end{equation}
The perturbation of the spatial metric in Fourier space becomes $h_{ij} = -2 \left( \Psi \delta_{ij} + k_i k_j E \right)$, its trace being $h = \delta^{ij} h_{ij}$. The continuity and Euler equations in this gauge read
\begin{align} \label{eq: nonadiabatic continuity euler sync}
   \delta'&=-(1+w)\left(\theta + \frac{h'}{2}\right)+3{\cal H}
   w\delta + 9{\cal H}^2c_a^2(1+w)\frac{\theta}{k^2} ,\\[5pt]
   \theta' &= -{\cal H}\theta , \label{eq: nonadiabatic euler sync}
\end{align}
respectively. Note that there exists a choice of initial conditions in the synchronous gauge which simplifies the treatment considerably. Indeed, if we choose $\theta|_\text{initial} = 0$, then equation \eqref{eq: nonadiabatic euler sync} implies that $\theta=0$ at all times. This in turn greatly simplifies the continuity equation \eqref{eq: nonadiabatic continuity euler sync}.\\[5pt]
\indent Summarizing, our general models will modify the cosmological evolution with respect to $\Lambda$CDM in two ways. On the one hand, the background quantities will have a different evolution from standard cosmology. In particular, they  modify the total dark sector equation of state and induce a non-vanishing adiabatic sound speed. On the other hand, due to their non-adiabaticity, they will change the structure of the perturbations equations
by including a new term in the continuity equation that is just proportional to $c_a^2$.


\section{Quadratic potential}\label{section: Quadratic potential}
\noindent To gain insight into the dynamical behavior of the proposed models, in this section we analyze the specific example of a quadratic potential. More in particular, we will consider
\begin{equation}\label{eq: quadratic potential}
    V(\psi) = \frac12 \xi \psi^2 + V_0 ,
\end{equation}
with $\xi$ a constant with dimensions of mass squared allowed to take both positive and negative values (implying, respectively, a valley or a hilltop potential), while $V_0$ represents the vertical offset. We can understand the $\xi\to0$ limit as the $\Lambda$CDM limit of the model, as it implies a constant potential. The background equations \eqref{eq: dpsi/da}, \eqref{eq: drhoc/da}, and \eqref{eq: Friedmann final} become
\begin{align}
    \frac{\d\psi}{\d a} &= \frac1{aH} \sqrt{2\left[ \beta \left( \frac12 \xi \psi^2 + V_0 \right) + \gamma \right]} , \label{eq: dpsi/da quadratic}\\[5pt]
    \frac{\d\rho_c}{\d a} &= - \frac3a \rho_c - (1-\beta) \xi \psi \frac{\d \psi}{\d a} , \label{eq: drhoc/da quadratic} \\[5pt]
    H^2 &= \frac{\kappa^2}{3} \Big[ \rho_\c + (1-\beta)\frac12 \xi \psi^2 + (1-\beta) V_0 - \gamma +\rho_{\B0}\, a^{-3} + \rho_{\R0}\, a^{-4} \Big] \label{eq: Friedmann quadratic} .
\end{align}
Let us introduce the normalized energy densities
\begin{equation}\label{eq: definition orhoc}
    \orho_i = \frac{\rho_i}{\crit} 
\end{equation}
for $i=\{\c,\lambda,\B,\R\}$, and where $\crit = 3H_0^2/\kappa^2$ is the critical energy density. Thus, the corresponding density parameters are just the present values
\begin{equation}
    \Omega_i = \bar\rho_{i0} .
\end{equation}
In particular, the vacuum energy density parameter in our chosen variables reads
\begin{equation}
    \Omega_\lambda = \frac12 \frac{(1-\beta)\xi}{\crit} \psi_0^2 + \frac{(1-\beta)V_0-\gamma}{\crit} ,
\end{equation}
with $\psi_0$ the value today. This helps us to rewrite the Friedmann equation \eqref{eq: Friedmann quadratic} as
\begin{equation}\label{eq: Friedmann quadratic clean}
    H^2 = H_0^2 \left[ \orho_\c +\frac12 \frac{(1-\beta)\xi}{\crit} \left(\psi^2-\psi_0^2\right) + \Omega_\lambda +  \Omega_\B \, a^{-3} + \Omega_\R \, a^{-4} \right] .
\end{equation}
Note that the cosmic sum rule
\begin{equation}\label{eq: cosmic sum rule}
    1 = \Omega_\c + \Omega_\lambda + \Omega_\B + \Omega_\R ,
\end{equation}
is satisfied, so that we can determine $\Omega_\lambda$ in terms of the other density parameters. In the following we shall study in detail the $\beta=0$ models for simplicity.

\subsection{$\beta=0$ models}
\noindent Setting $\beta=0$ yields great simplifications. It is possible to bring the background system \eqref{eq: dpsi/da quadratic}, \eqref{eq: drhoc/da quadratic}, and \eqref{eq: Friedmann quadratic clean} into the form
\begin{subequations}\label{eq: beta0 system}
    \begin{align}
    \frac{\d\opsi}{\d a} &= \frac1{aE} , \label{eq: constraint beta0}\\[5pt]
    \frac{\d\orho_\c}{\d a} &= - \frac3a \orho_\c - \oxi\,\opsi\, \frac{\d \opsi}{\d a} , \label{eq: conservation beta0} \\[5pt]
    E^2 &= \orho_\c + \frac12 \oxi \left(\opsi^2-\opsi_0^2\right) + \Omega_\lambda +  \Omega_\B \, a^{-3} + \Omega_\R \, a^{-4} \label{eq: Friedmann beta0} ,
\end{align}
\end{subequations}
where $E=H/H_0$, $\orho_\c$ is given by \eqref{eq: definition orhoc}, and we have introduced the dimensionless field and coupling
\begin{align}
    \opsi &= \frac{H_0\psi}{\sqrt{2\gamma}} ,\\[5pt]
    \oxi &= \frac{2\gamma \xi}{H_0^2\crit} .
\end{align}
With these redefinitions, we observe that the new system \eqref{eq: beta0 system} depends only on three parameters (besides the usual $\Omega_\B$ and $\Omega_\R$). These are the two initial conditions for \eqref{eq: constraint beta0} and \eqref{eq: conservation beta0}, respectively $\opsi_0$ and $\Omega_\c$ (indeed, $\orho_{\c0} = \Omega_\c$) and the potential parameter $\oxi$.

The interacting kernel \eqref{eq: projected kernel Q} in these models becomes
\begin{equation}\label{eq: Q in beta0}
    Q =  -\oxi\,\opsi\,H_0\crit \,aE\, \frac{\d\opsi}{\d a} = -\oxi\,\opsi\, H_0\crit ,
\end{equation}
where in the final equality we have used the field equation \eqref{eq: constraint beta0}. From the above expression it is clear that energy transfer will reverse the moment the field crosses $\opsi=0$, as the sign of $Q$ will switch. Moreover, note that the sign of the constant $\oxi$ determines whether dark matter will receive or give energy to the vacuum when $\opsi<0$ or $\opsi>0$.

\medskip
\noindent\textbf{Cosmological parameters.} In order to show the potential of these models in describing a dynamical dark energy in the light of the recent DESI BAO results, we will carry out a simplified comparison between TDiff and CPL. A detailed statistical analysis of the data will be performed elsewhere. In order to numerically solve system \eqref{eq: beta0 system} we will take, in the visible sector, the Big Bang nucleosynthesis value $\Omega_\B h^2 = 0.02205$ (where $H_0 = 100 h \ \text{km}\,\text{s}^{-1}\,\text{Mpc}^{-1}$), together with $\Omega_\gamma h^2 = 2.47\times10^{-5}$ and $N_\text{eff} = 3.044$ \cite{ParticleDataGroup:2024cfk}, which result in $\Omega_\R h^2 = 4.18\times10^{-5}$. In the dark sector, for $\Lambda$CDM and CPL we will consider the best-fit values found in the DES-Dovekie + CMB + BAO analysis \cite{DES:2025sig}, and for the TDiff model some example parameters. All in all, then, the parameters we shall employ are:
\begin{subequations}\label{eq: parameters beta0}
    \begin{align}
        &\text{$\Lambda$CDM:} \ \{ h = 0.6814 ,\ \ \Omega_\c = 0.257 \} , \label{eq: parameters beta0 LCDM}\\[5pt]
        &\text{CPL:}\ \{ h = 0.6747 ,\ \ \Omega_\c = 0.265, \ \ w_0 = -0.803 ,\ \ w_a = -0.72 \} , \label{eq: parameters beta0 CPL}\\[5pt]
        &\text{TDiff:} \ \{ h= 0.6775 ,\ \ \Omega_\c = 0.414 ,\ \ \oxi = -2.54 ,\ \ \opsi_0 = 0.435 \} \label{eq: parameters beta0 TDiff} .
    \end{align}
\end{subequations}
Notice that in the TDiff model the interacting dark matter does not necessarily scale as $a^{-3}$, and hence the present abundance $\Omega_\c$ can differ from $\Lambda$CDM and CPL. Moreover, notice also that the TDiff model has the same number of free parameters for the dark sector as CPL.

\begin{figure*}[t]
    \centering
    {\includegraphics[width=0.496\linewidth]{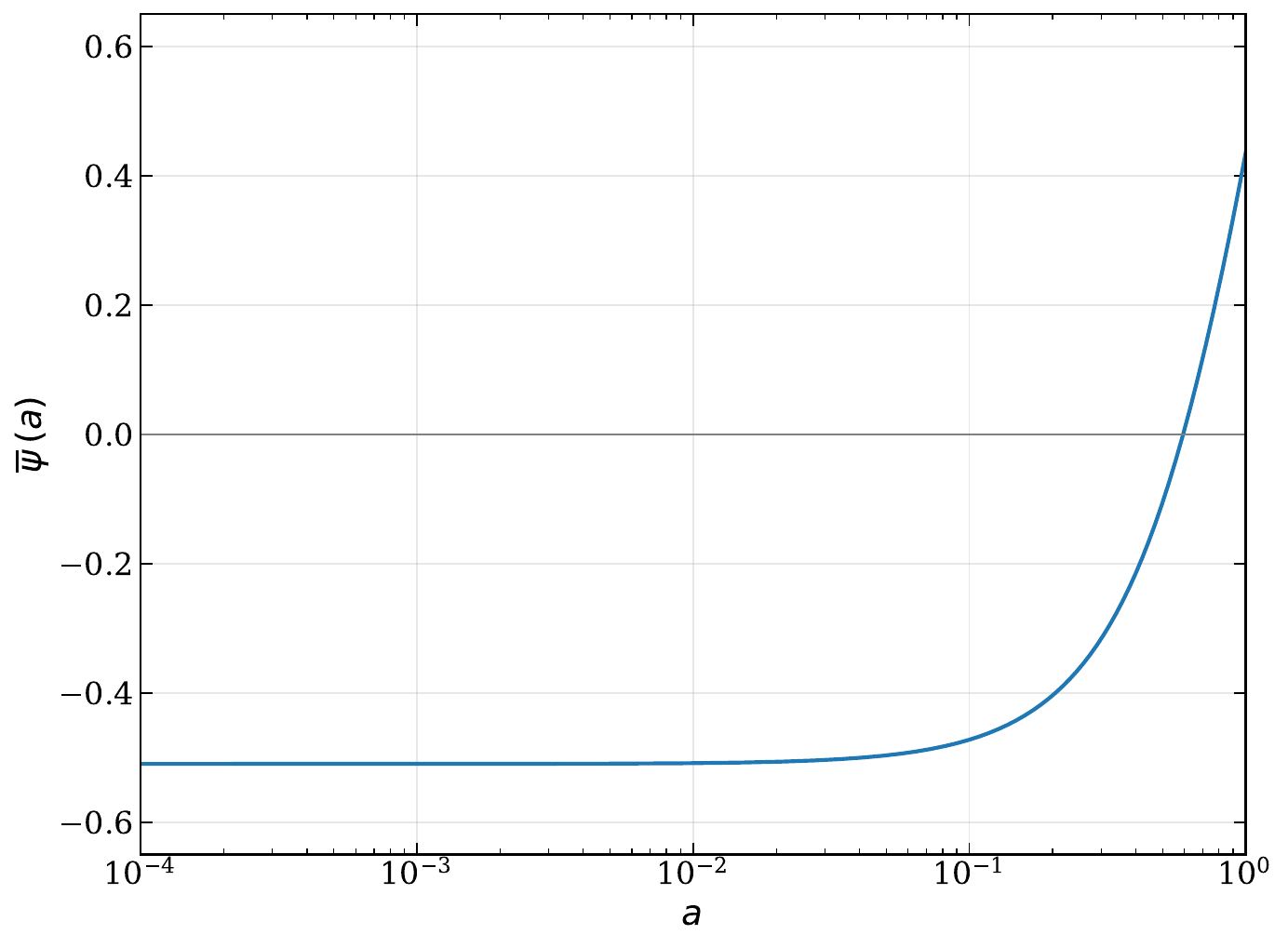} \includegraphics[width=0.496\linewidth]{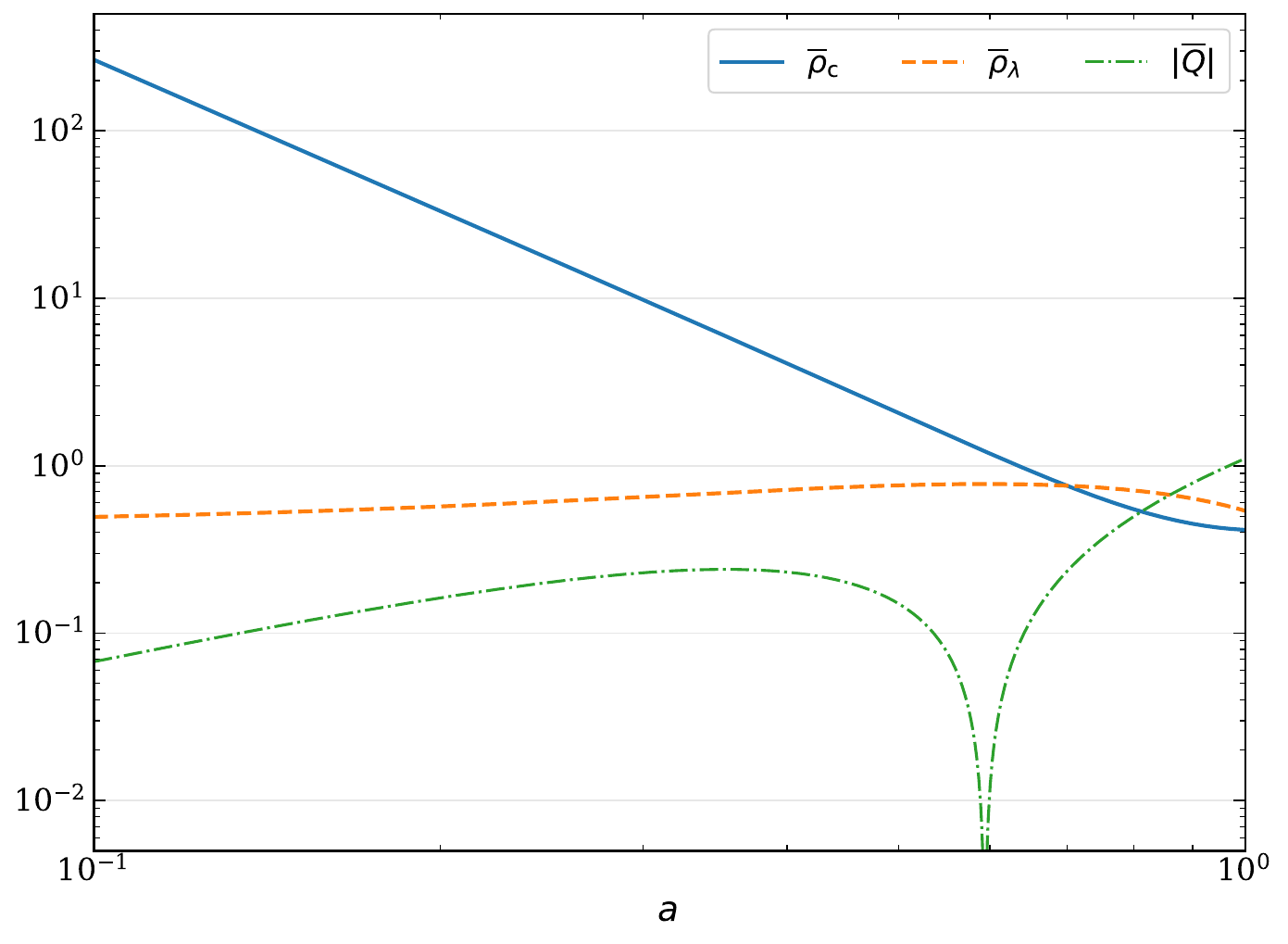}}
    \caption{\footnotesize On the left panel we show the evolution of the dimensionless field $\opsi$ for the TDiff model with parameters \eqref{eq: parameters beta0 TDiff}. The transition from negative to positive values occurs at around $a\simeq 0.594$. On the right panel, we depict the
    evolution of the normalized energy densities $\orho_\c$ and $\orho_\lambda$ in the TDiff model \eqref{eq: parameters beta0 TDiff}, together with the (absolute value of the) normalized interacting kernel $\bar{Q}$. We  see that at early times
    $\bar \rho_c$ scales as cold dark matter and $\bar\rho_\lambda$ as a cosmological constant, whereas they deviate from the $\Lambda$CDM behavior at late times.}
    \label{fig: beta0_psi}
\end{figure*}

\begin{figure}[pos=t]
    \centering
    \includegraphics[width=\linewidth]{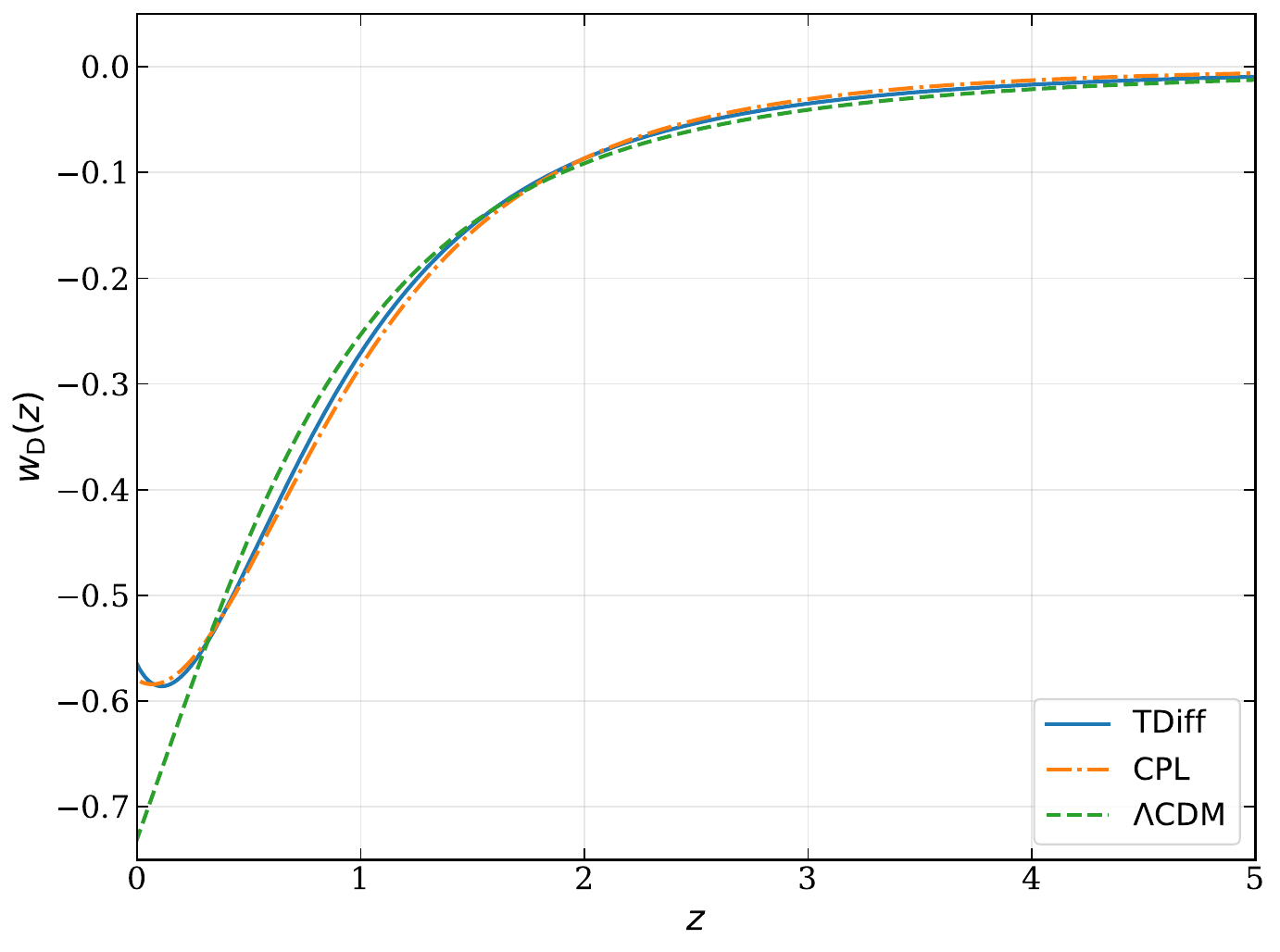}
    \caption{\footnotesize Evolution of the complete dark sector's equation of state for the models and parameters presented in \eqref{eq: parameters beta0}. The solid blue line represents the unified $\beta=0$ TDiff model \eqref{eq: DS EoS}, the dashed green line is a $\Lambda$CDM dark sector \eqref{eq: wD LCDM} and the dot-dashed orange line is a CPL dark sector \eqref{eq: wD CPL}.}
    \label{fig: beta0_w}
\end{figure}

\medskip
\noindent\textbf{Cosmological background.} We present in the left panel of Figure \ref{fig: beta0_psi} the evolution of the dimensionless field $\opsi$. It initially takes on negative values and eventually crosses zero at around $a\simeq 0.594$, triggering a sign flip in the interacting kernel \eqref{eq: Q in beta0} with consequences which will be discussed in the following. Interestingly, we also notice that the field is approximately constant at early times (until about $a\sim 10^{-2}$). This means that the vacuum energy contribution will also be approximately constant itself as follows from its definition in \eqref{eq: rho decomposition}, and the TDiff model effectively behaves as $\Lambda$CDM in the early universe.
We may verify this expectation by explicitly studying the evolution of the dust and vacuum energies. The right panel of Figure \ref{fig: beta0_psi} shows the evolution of the normalized energy densities $\orho_\c$ and $\orho_\lambda$, together with the normalized interacting kernel
\begin{equation}
    \bar{Q} = \frac{Q}{H\crit} ,
\end{equation}
which measures the energy transfer in units of the critical energy density and the instantaneous Hubble rate. One may see that in the early universe we have the expected scalings $\rho_\c \sim a^{-3}$ and $\rho_\lambda\sim\text{const.}$ (and, in particular, the dust contribution dominates the dark sector's energy), but they eventually deviate from such behavior as a result of the interaction. Indeed, the interaction is so inefficient in the early universe that it is effectively switched off, making the dust and vacuum components behave as in standard $\Lambda$CDM. As the universe evolves, the interaction gains importance in relation to the values of the dust and vacuum energies, making the latter appreciably grow.
Eventually, however, the kernel switches sign as the field crosses zero and the behavior is reversed: the vacuum decays and gives off energy to the dust component. The effect of this reversal is most clearly appreciated when one considers the evolution of the complete dark sector's EoS, shown in Figure \ref{fig: beta0_w} for the models and parameters presented in \eqref{eq: parameters beta0}. We stress that we are studying the EoS of the complete dark sector, i.e. we are using
\begin{align}
    w_{\D,\Lambda\text{CDM}}(a) &= \frac{-1}{1 + \frac{\Omega_\c}{\Omega_\Lambda} a^{-3}} , \label{eq: wD LCDM} \\[5pt]
    w_{\D,\text{CPL}}(a) &= \frac{\big[ w_0 + w_a(1-a) \big] a^{-3(1+w_0+w_a)} e^{3w_a(a-1)}}{\frac{\Omega_\c}{\Omega_\DE} \,a^{-3} + a^{-3(1+w_0+w_a)} e^{3w_a(a-1)}} , \label{eq: wD CPL}
\end{align}
where $\Omega_\Lambda$ represents the cosmological constant abundance in $\Lambda$CDM, and $\Omega_\DE$ represents the DE abundance in CPL (both may be found from the respective cosmic sum rules). The most remarkable feature observed in Figure \ref{fig: beta0_w} is the fact that the unified TDiff dark sector is able to reproduce the behavior of the CPL dark sector at all redshifts. The late-time weakening of the dark energy found with the CPL parametrization can be understood in the TDiff model, using the interacting decomposition, as a result of the sign-switching of the interacting kernel.

\begin{figure}[pos=t]
    \centering
    \includegraphics[width=\linewidth]{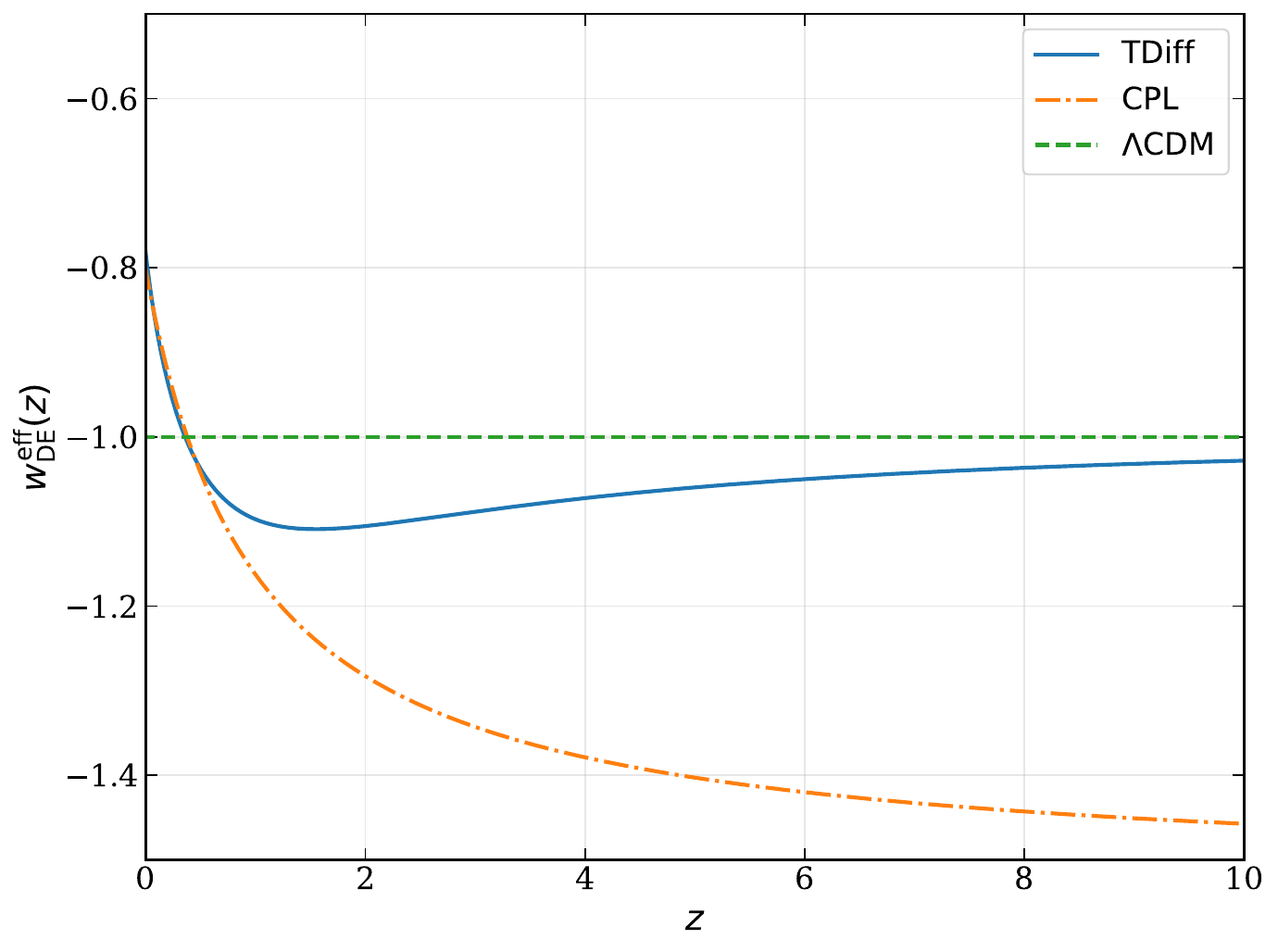}
    \caption{\footnotesize Evolution in redshift of the effective DE's equation of state for the models and parameters presented in \eqref{eq: parameters beta0}. The solid blue line represents the $\beta=0$ TDiff model \eqref{wDEeff}, the dot-dashed orange line corresponds to the CPL parametrization $w_\text{CPL}(z) = w_0 + w_a \frac{z}{1+z}$, and the dashed green line is simply the cosmological constant from $\Lambda$CDM, $w_\Lambda = -1$.}
    \label{fig: beta0_eff}
\end{figure}

\medskip
\noindent\textbf{A non-interacting decomposition of the dark sector.} Inspired by the above results, we can attempt a different partition to the one we have thus far considered. Instead of (formally) divvying up the unified dark sector as the sum of two interacting perfect fluids with constant EoS parameters ($w_\c = 0$, $w_\lambda=0$), let us partition it as the sum of two \textit{non}-interacting perfect fluids (hence separately conserved) as follows:
\begin{equation}\label{eq: effective decomposition}
    \rho_\D = \rho_\DM^\eff + \rho_\DE^\eff ,
\end{equation}
i.e. as the sum of effective DM and DE pieces. We will ask that $w_\DM^\eff = 0$, thus imposing a standard decay on the effective DM piece
\begin{equation}\label{eq: effective rhoc}
    \rho_\DM^\eff = \rho_{\DM0}^{\eff} \, a^{-3} ,
\end{equation}
while attributing all other evolution to the effective DE component. It is possible to numerically compute $\rho_{\DM0}^{\eff}$ taking advantage of the fact that the dust component dominates the dark sector in the early universe, as we have just seen. Evaluating the early-time behavior
\begin{equation}\label{eq: effective rhoc0}
	\rho_\D(a_\text{early}) \simeq \rho_{\DM0}^{\eff} \, a_\text{early}^{-3} \,\implies\, \rho_{\DM0}^{\eff} \simeq \rho_\D(a_\text{early}) \, a_\text{early}^3
\end{equation}
with $a_\text{early} \lsim 10^{-2}$, yields
\begin{equation}\label{eq: effective Omegac}
    \Omega_\DM^\eff = \frac{\rho_{\DM0}^{\eff}}{\crit} =0.266 .
\end{equation}
This parameter admits the usual interpretation of DM abundance and its value may be compared with those in $\Lambda$CDM and CPL as presented in \eqref{eq: parameters beta0 LCDM} and \eqref{eq: parameters beta0 CPL}, showing overall consistency. On another note, the requirement $w_c^\eff=0$ implies that the total dark sector pressure will be purely that of the effective DE piece, $p_\lambda = p_\D = p_\DE^\eff$. The price to pay in this non-interacting decomposition is that the EoS of the remaining component $w_\DE^\eff$ will not be constant and will instead present evolution, but it is as simple as
\begin{equation}
    w_\DE^\eff = \frac{p_\DE^\eff}{\rho_\DE^\eff} = \frac{p_\D}{\rho_\D - \rho_\DM^\eff} .\label{wDEeff}
\end{equation}
The evolution of the effective DE EoS is shown in Figure \ref{fig: beta0_eff} for the TDiff model as well as for the CPL parametrization $w_{\DE,\text{CPL}}(a) = w_0 + w_a (1-a)$ and $\Lambda$CDM model $w_\Lambda = -1$ using the parameter values presented in \eqref{eq: parameters beta0}. Note that, even though the unified TDiff dark sector never crosses the phantom divide, in the non-interacting decomposition the effective dark energy eventually does. Moreover, even though the TDiff model reproduces the total dark sector CPL equation of state at all redshifts, the effective dark energy EoS differs significantly from CPL at high redshifts. In particular, it approaches $w_\DE^\eff \to -1$ at early times, which is consistent with our findings that the model effectively behaves as $\Lambda$CDM early on.

\begin{figure*}[t]
    \centering
   { \includegraphics[width=0.496\linewidth]{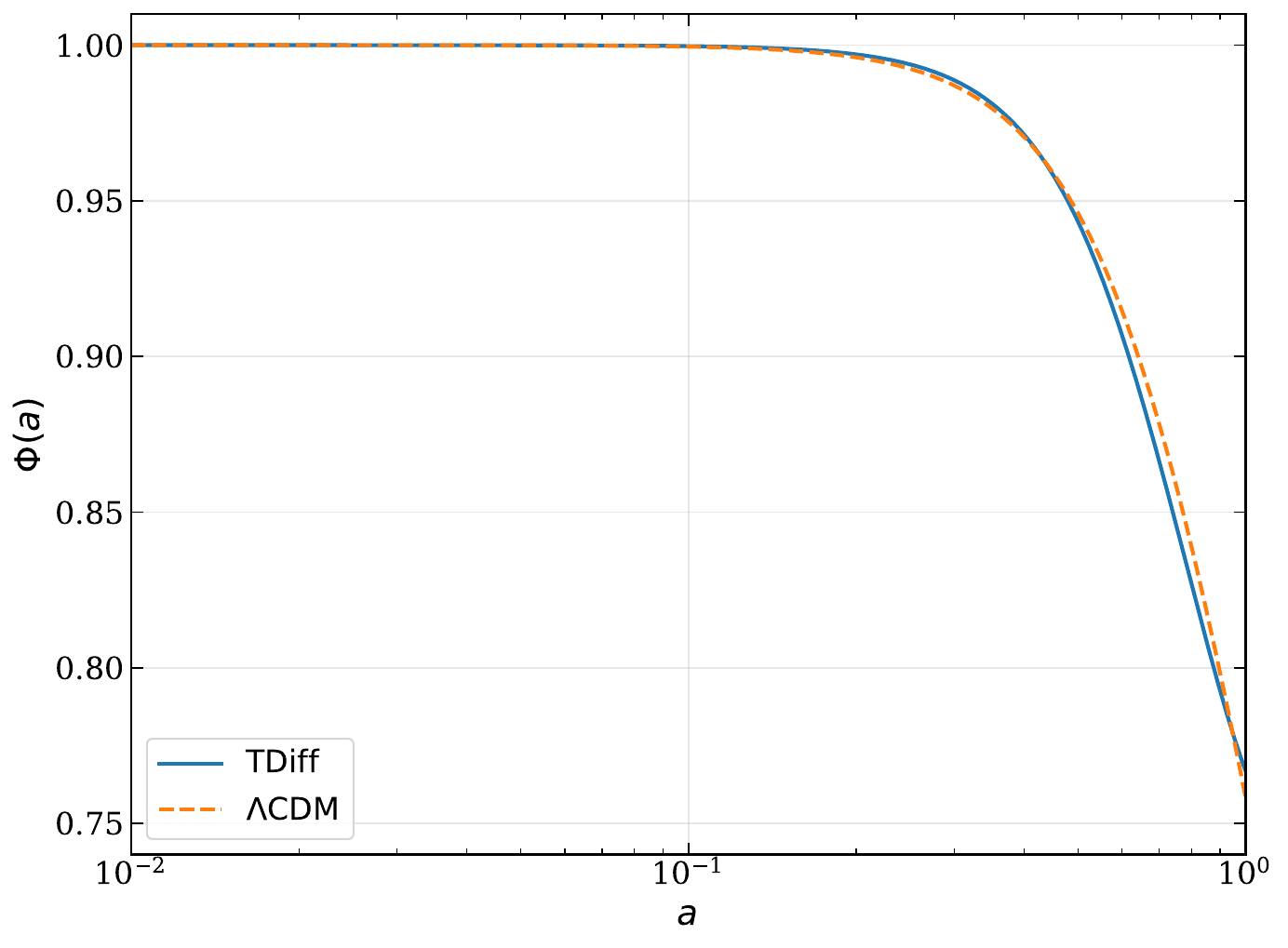}
    \includegraphics[width=0.496\linewidth]{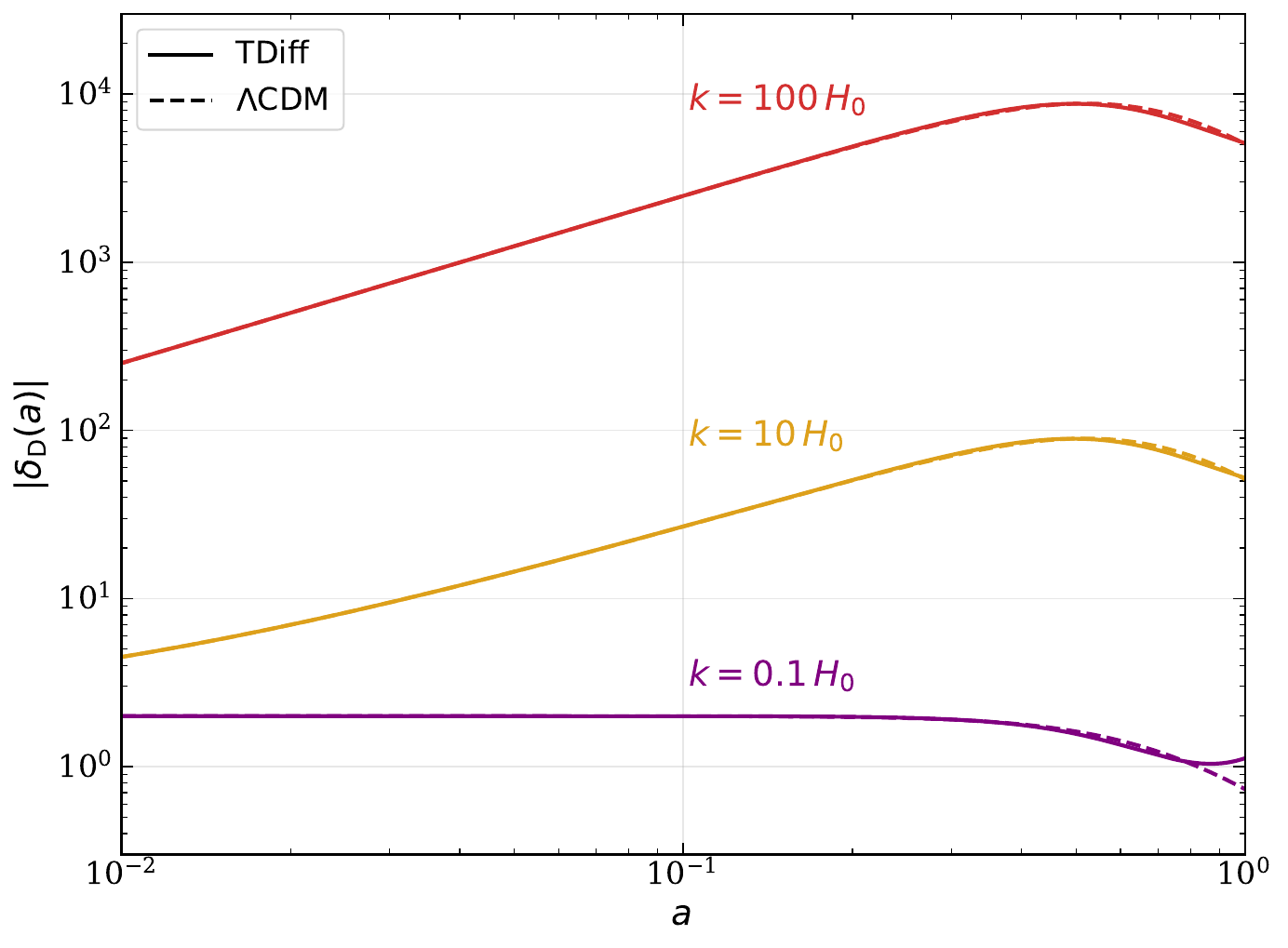}}
    \caption{\footnotesize For a dark-sector-only ($\Omega_\B=\Omega_\R=0$) TDiff model \eqref{eq: parameters beta0 TDiff} and $\Lambda$CDM model with $\Omega_\c = 0.268$, we show on the left panel the evolution of the gravitational potential $\Phi$  and, on the right panel, the evolution of the (absolute value of the) total dark sector density contrast for three different scales.}
    \label{fig: beta0_Phi}
\end{figure*}

\medskip
\noindent\textbf{Cosmological perturbations.} Here we will focus only on the evolution of the dark sector perturbations, so that for simplicity we will neglect the contributions from baryons and radiation. 
In this approximation it suffices to find the evolution of the metric perturbation $\Phi$ to determine that of $\delta$ and $\theta$. To see how, we begin by writing the (conformal time version of the) background Friedmann and acceleration equations:
\begin{align}
    \H^2 &= \frac{\kappa^2}3 a^2\rho_0 , \label{eq: conformal Friedmann} \\[5pt]
    \H' &= -\frac12 \H^2(1+3w) \label{eq: conformal acceleration} ,
\end{align}
where we are assuming that only the dark fluid enters at the background level, and for the current calculations we will adopt back the notation of a prime for conformal time derivatives and a subindex ``0'' for background quantities. Using \eqref{eq: conformal Friedmann} and $\theta=-k^2v$ on the first two perturbed Einstein equations \eqref{eq: einstein with delta rho} and \eqref{eq: einstein with v}, respectively, gives
\begin{align}
    \delta &= -\frac23 \left[ \frac{k^2}{\H^2} \Phi + 3 \left( \Phi + \frac{\Phi'}\H \right) \right] ,\label{eq: delta and Phi} \\[5pt]
    \theta &= \frac23 \frac{k^2}{\H^2} \frac{\H \Phi + \Phi'}{1+w} \label{eq: theta and Phi}
\end{align}
in Fourier space. This shows that the evolution of the density contrast $\delta$ and velocity divergence $\theta$ is purely determined by that of the gravitational potential $\Phi$. Thus, one just needs to find an equation for it. This is achieved by using \eqref{eq: theta and Phi} to write the pressure perturbation \eqref{eq: short pressure perturbation} as
\begin{equation}
    \delta p = - 2 c_a^2 \left( \Phi + \frac{\Phi'}\H \right) \rho_0 .
\end{equation}
Substituting this result in the remaining perturbed Einstein equation \eqref{eq: einstein with delta p} and using the conformal Friedmann and acceleration equations \eqref{eq: conformal Friedmann} and \eqref{eq: conformal acceleration} gives
\begin{equation}
    \Phi'' + 3\H \left(1+c_a^2\right) \Phi' + 3\H^2 \left(c_a^2-w\right) \Phi = 0 ,
\end{equation}
or equivalently
\begin{equation}\label{eq: Phi ODE}
    \frac{\d^2\Phi}{\d a^2} + \frac{1}{2a}\left(7-3w+6c_a^2\right) \frac{\d\Phi}{\d a} + \frac{3}{a^2} \left(c_a^2-w\right) \Phi = 0 .
\end{equation}
Note that, as opposed to \eqref{eq: delta and Phi} and \eqref{eq: theta and Phi}, the wavenumber $k$ does not appear in the evolution equation for $\Phi$, which means that the gravitational potential will evolve identically on all scales. For the purposes of the present work this simplified cosmological picture will be enough, and any behavior we extract from it may be understood as the leading behavior in the periods of the dark fluid domination. Nonetheless, a more detailed analysis should take into account the presence of the rest of the energy content of the universe, and this will be considered in future work. We numerically solved \eqref{eq: Phi ODE} and plot on the left of Figure \ref{fig: beta0_Phi} the evolution of the gravitational potential $\Phi$ for a dark-sector-only ($\Omega_\B=\Omega_\R=0$) TDiff model \eqref{eq: parameters beta0 TDiff} and $\Lambda$CDM model with $\Omega_\c = 0.268$. We see that the early-time behavior of $\Phi$ in the TDiff model agrees with $\Lambda$CDM, however they differ at late times, which could affect the integrated Sachs-Wolfe effect of the CMB anisotropies. The plot on the right of Figure \ref{fig: beta0_Phi} then shows the evolution of the dark sector density contrast for three different scales. We observe that the evolution of scales which are subhorizon today ($k\gg H_0$) is practically identical to that of $\Lambda$CDM, while the evolution of scales which are superhorizon today ($k\ll H_0$) starts differing appreciably only at late time. In this manner, we expect that the $\beta=0$ TDiff model has a very similar history of structure formation to $\Lambda$CDM.

\section{Conclusions}\label{section: Conclusions}
\noindent In this work we have introduced a novel class of field-theory models which provide a unified description of the dark sector as a perfect fluid with vanishing speed of sound. 
Such a description naturally arises in the simplest TDiff scalar field theory, where arbitrary functions of the metric determinant mediate the coupling to gravity. These coupling functions are then fixed when focusing attention to the silent case, i.e. that leading to a perfect fluid with a vanishing speed of sound.

As is known, when considering TDiff scalar theories, Diff symmetry can be restored introducing new fields. In our case, this yields
a parametrized family of two-field scalar Diff theories. In fact, the class of theories with
a vanishing sound speed are found to be equivalent to a mimetic framework.
We have also verified that the resulting unified dark sectors never cross the phantom line, they generally describe non-adiabatic perfect fluids and admit the interpretation of interacting vacuum models. Moreover, for this interpretation the underlying fundamental description provides a concrete theoretical expression of the interacting kernel. 
In particular, the $\Lambda$CDM dark sector is recovered as a particular case in which the interaction vanishes and 
the unified fluid is adiabatic.

After the general treatment, we have focused on a cosmological context and studied the quadratic potential as a concrete example, showing the capabilities of these models in describing the dark sector of the universe in light of the recent DESI results hinting at a dynamical dark energy. To this end, we have studied in detail a simple realization of our TDiff models with the same number of free parameters as the CPL parametrization. At the background level, the unified dark sector is $\Lambda$CDM-like at early times but the evolution proceeds differently later on. Remarkably, we have found it is possible for the complete dark sector equation of state of the TDiff model to reproduce CPL at all redshifts. However, when assuming a non-interacting decomposition of our unified fluid, we have shown that the effective dark energy equation of state differs, especially at early times: while CPL falls to phantom, the TDiff model considered tends to a cosmological constant in the early universe. At the perturbative level, we have carried out a simplified (dark-sector-only) analysis, and compared the TDiff model with $\Lambda$CDM. On the one hand, the evolution of the gravitational potential has been found to be identical to that of $\Lambda$CDM at high redshift, but the two models deviate at lower redshift and this could affect the integrated Sachs-Wolfe effect. On the other hand, the density contrast of the complete dark sector in the TDiff model evolves almost identically to $\Lambda$CDM at early times for all scales, while at late times we find small differences only in long wavelength modes. For this reason, we expect a structure formation history similar to that of $\Lambda$CDM.

It is worth emphasizing that the class of models considered in this work is quite broad. Indeed, one could explore different forms for the potential, different parameter choices, and in general more complete realizations beyond the simple example presented in this work, which already illustrates the potential of the framework. Our findings therefore open up a number of interesting avenues for future investigation. Firstly, a detailed statistical analysis confronting TDiff models with CMB, supernovae, and BAO data is necessary to assess their observational viability, and will be carried out in future work. Secondly, the analysis of CMB anisotropies and the non-linear regime of structure formation could provide further means of distinguishing TDiff models from CPL and $\Lambda$CDM. In this context, \cite{Kou:2025yfr} investigated non-linear structure formation using a concrete spherical collapse model, and found it challenging to significantly discriminate the unified fluid from CPL. For the models here discussed, however, the existence of an underlying field theory provides, in principle, access to the full non-linear dynamics, making it possible to study structure formation without the need to prescribe a collapse model. More broadly, our results illustrate that TDiff theories provide a well-motivated theoretical framework for constructing unified and phenomenologically viable descriptions of the dark sector.

\section*{Acknowledgements}
\noindent This work has been supported by the MICIN (Spain) Project No. PID2022-138263NB-I00 funded by MICIU/AEI/10.13039/ 501100011033 and by ERDF/EU.
DJG acknowledges support from the Comunidad de Madrid under predoctoral contract PIPF-2023/TEC29931. JdCP was supported by the projects AST22\_00001 and AST22\_8.4\_SR funded by the European Union and by the Regional Government of Andaluc{\'i}a.

\bibliographystyle{elsarticle-num}
\bibliography{references}

\end{document}